\documentclass[prb,twocolumn,showpacs,amsmath,amssymb,superscriptaddress,bibnotes]{revtex4-2}
\usepackage{graphicx}
\usepackage{color}
\usepackage{braket}
\usepackage{amsmath}
\usepackage{amsfonts}
\usepackage[caption=false]{subfig}

\usepackage{amsmath,amssymb,mathrsfs}
\usepackage{bbm}
\usepackage[dvipsnames]{xcolor}
\usepackage{appendix}
\usepackage{verbatim}
\usepackage{bbold}

\date{\today}

\begin{document}
	
	\title{Fractional second-order topological insulator from a three-dimensional \\coupled-wires construction}

	\author{Katharina Laubscher}
	\thanks{These authors contributed equally to this work.}
	\affiliation{Department of Physics, University of Basel, Klingelbergstrasse 82, CH-4056 Basel, Switzerland}
	\author{Pim Keizer}	
	\thanks{These authors contributed equally to this work.}
	\affiliation{Department of Applied Physics, Eindhoven University of Technology, 5600 MB Eindhoven, The Netherlands}
	\author{Jelena Klinovaja}
	\affiliation{Department of Physics, University of Basel, Klingelbergstrasse 82, CH-4056 Basel, Switzerland}
	
	\date{\today}
	
	\begin{abstract}
		We construct a three-dimensional second-order topological insulator with gapless helical hinge states from an array of weakly tunnel-coupled Rashba nanowires. For suitably chosen interwire tunnelings, we demonstrate that the system has a fully gapped bulk as well as fully gapped surfaces, but hosts a Kramers pair of gapless helical hinge states propagating along a path of hinges that is determined by the hierarchy of interwire tunnelings and the boundary termination of the system. Furthermore, the coupled-wires approach allows us to incorporate electron-electron interactions into our description. At suitable filling factors of the individual wires, we show that sufficiently strong electron-electron interactions can drive the system into a fractional second-order topological insulator phase with hinge states carrying only a fraction $e/p$ of the electronic charge $e$ for an odd integer $p$. %Our toy model adds to the general understanding of second-order topological insulators and, in particular, provides an explicit analytical description of an exotic interaction-driven higher-order topological phase.
	\end{abstract}
	
	\maketitle

	\section{\label{sec:Introduction} Introduction}
	Recently, the concept of so-called higher-order topological insulators (HOTIs) has significantly enriched the existing classification of topological phases of \mbox{matter~\cite{Benalcazar2017,Benalcazar2017b,Imhof2018,Song2017,Peng2017,Schindler2018,Geier2018,Langbehn2017}}. While a conventional $d$-dimensional topological insulator (TI) exhibits gapless boundary states at its $(d-1)$-dimensional boundaries, a $d$-dimensional $n$th-order TI hosts gapless states at its $(d-n)$-dimensional boundaries. Of particular interest in this context are second-order TIs, which host topologically protected zero-energy corner states (gapless hinge states) in two (three) dimensions~\cite{Miert2018,Ezawa2018a,Ezawa2018b,Khalaf2018,Hsu2018,Okugawa2019,Calugaru2019,Nag2019,Volpez2019,Sheng2019,Benalcazar2019,Franca2019,Zhang2019,Plekhanov2020,Laubscher2020b,Yoon2020,Tiwari2020,Zhang2020b,Roy2019,Plekhanov2019,Li2021,Ghosh2021}. Experiments have reported signatures of second-order TI phases in a few materials~\cite{Schindler2018b,Noguchi2021,Shumiya2022,Aggarwal2021,Choi2020,Kononov2020} as well as in various artificial structures based on, e.g., mechanical, phononic, photonic, or electrical systems~\cite{Garcia2018,Xue2019,Ni2019,Mittal2018,Hassan2018,Xie2018,Zhang2020,Noh2018}.
	
While the original theory of HOTIs builds on single-particle band structure considerations, it is interesting to ask whether there are exotic interaction-driven HOTI phases that do not fit into this conventional picture. While this question has by now been answered affirmatively~\cite{You2018a,You2018b,Laubscher2019,Laubscher2020,May-Mann2022,Hackenbroich2021,Zhang2022,Zhang2022b,Li2022}, concrete toy models for strongly interacting HOTI phases are still extremely rare since analytical tools to study interacting systems in more than one dimension are scarce. One way forward is offered by the so-called coupled-wires approach~\cite{Kane2002,Teo2014}, where a two-dimensional (2D) or three-dimensional (3D) system is modeled as an array of weakly coupled one-dimensional (1D) wires. This then allows for an analytical treatment of strong electron-electron interactions via the standard 1D bosonization formalism~\cite{Giamarchi2004}. Models of coupled wires have been used with high success to study a variety of exotic interacting first-order topological phases, including, for example, fractional quantum Hall states~\cite{Kane2002,Klinovaja2013,Teo2014,Klinovaja2014b,Klinovaja2015,Sagi2015a,Tam2021,Laubscher2021}, fractional TIs~\cite{Klinovaja2014a,Sagi2014,Neupert2014,Santos2015,Sagi2015b,Meng2015}, or interacting topological superconductors~\cite{Neupert2014,Sagi2017,Li2020}. As for HOTI phases, models for strongly interacting second-order topological superconductors in two dimensions~\cite{Laubscher2019,Laubscher2020,Zhang2022b} and certain classes of 3D second-order TIs protected by subsystem symmetries~\cite{May-Mann2022} have been brought forward. Nevertheless, a large variety of interacting HOTI phases do not yet have concrete realizations.
	
		\begin{figure}[b]
		\centering
		\includegraphics[width=0.8\columnwidth]{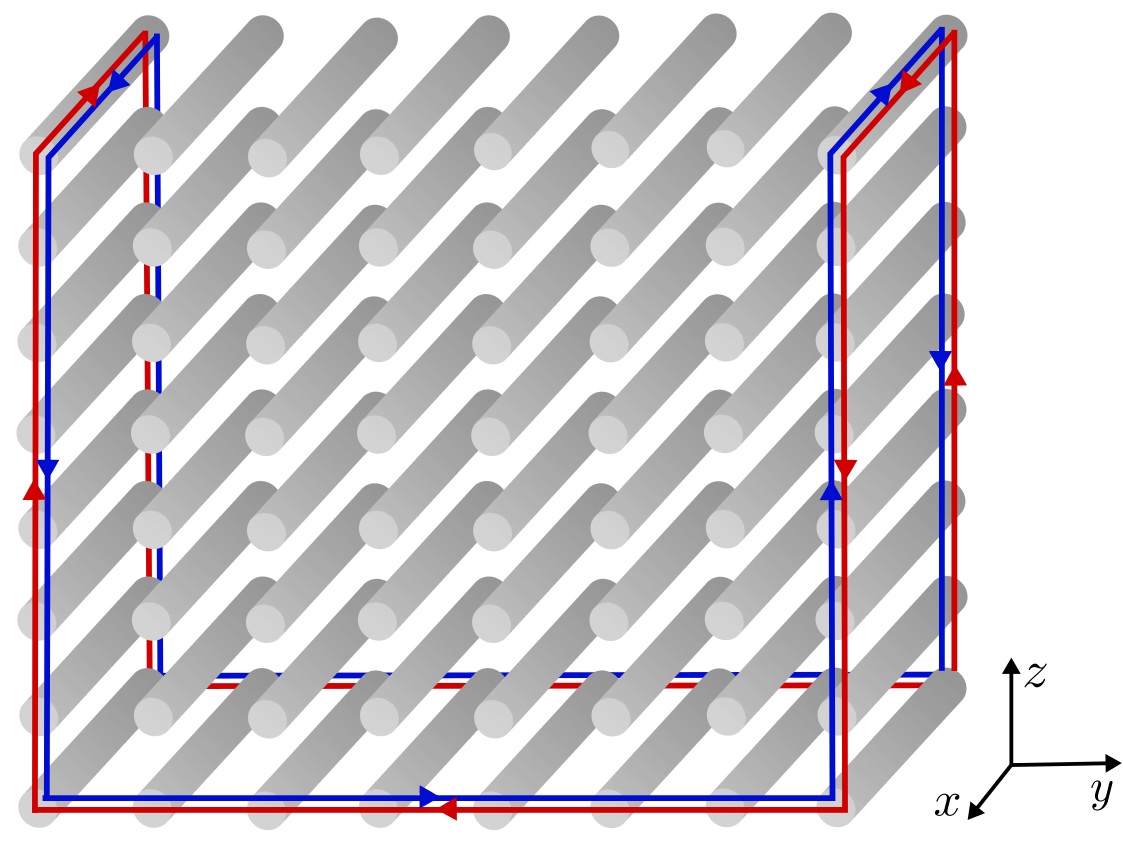}
		\caption{Sketch of the coupled-wires model. The starting point is a 3D array of uncoupled Rashba nanowires (gray), which are taken to be aligned along the $x$ axis. For suitable interwire couplings, the system is in a helical HOTI phase with a Kramers pair of gapless hinge states (shown in red and blue) propagating along a closed path of hinges.}
		\label{fig:Wires_uniform}
	\end{figure}
	
With this motivation, we construct a 3D coupled-wires model that is capable of realizing different second-order TI phases with gapless helical hinge states protected by time-reversal symmetry, see Fig.~\ref{fig:Wires_uniform}. For suitably chosen interwire couplings,  the non-interacting system has a fully gapped bulk as well as fully gapped surfaces but hosts a Kramers pair of gapless helical hinge states propagating along a path of hinges that is determined by the hierarchy of interwire tunnelings and the boundary termination of the system. Furthermore, at fractional fillings and for sufficiently strong electron-electron interactions, the system can enter an exotic fractionalized phase with gapless helical hinge states that carry only a fraction of the electronic charge $e$. All of the HOTIs constructed in this work should be seen as \emph{extrinsic} HOTIs in the language of Ref.~\cite{Geier2018}, meaning that the helical hinge states are protected by time-reversal symmetry as well as the minimum surface gap (rather than the 3D bulk gap). In this case, no additional crystalline symmetries are needed to protect the helical hinge states. As opposed to previous work in a similar direction~\cite{May-Mann2022}, where even the non-fractionalized phase required the presence of interaction terms in order to maintain the protecting subsystem symmetry, we obtain conventional hinge states with charge $e$ from only single-particle tunnelings between nearest-neighbor wires.
	
This paper is organized as follows. In Sec.~\ref{sec:Model}, we introduce our model of weakly tunnel-coupled Rashba nanowires. In Sec.~\ref{sec:HOTI}, we then show that this model exhibits a HOTI phase with gapless helical hinge states that propagate along a closed path of hinges of a finite 3D sample. We explain in detail how the path of the hinge states is determined by the hierarchy of interwire tunnelings and the boundary termination of the system, and provide some general insight into the construction of 3D HOTIs from lower-dimensional building blocks. In Sec.~\ref{sec:FractionalHOTI} we include electron-electron interactions into our description and show that, for sufficiently strong interactions, a fractional HOTI phase with fractionally charged hinge states can be realized. We conclude in Sec.~\ref{sec:conclusions}.
	
\section{Model}
\label{sec:Model}
\begin{figure}[tb]
	\centering
	\includegraphics[width =0.95\columnwidth]{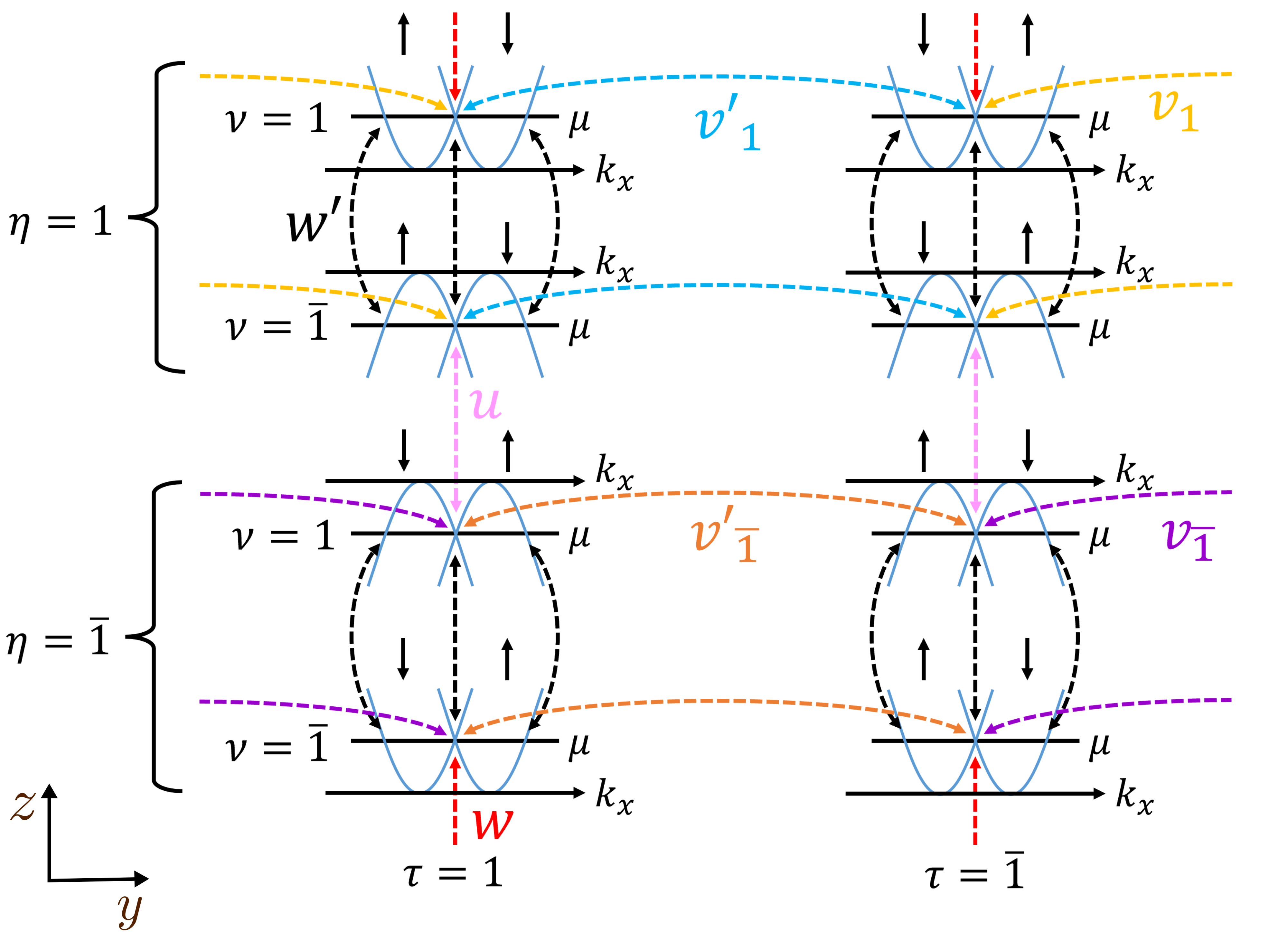}
	\caption{Schematic depiction of a single unit cell of the coupled-wires construction described in the main text. The spectrum of the uncoupled wires [see Eq.~(\ref{eq:Spectrum_continuum})] is represented by the blue parabolas. Wires with $\eta\nu = 1$ ($\eta\nu = \bar{1}$) have a positive (negative) effective mass and  wires with $\tau\nu = 1$ ($\tau\nu = \bar{1}$) have positive (negative) SOI. The chemical potential $\mu$ is tuned to the crossing point between spin-up and spin-down branches at $k_x = 0$ and the dashed lines represent the different interwire tunnelings [see Eqs.~(\ref{eq:Inter_z})-(\ref{eq:Intra_y})].}
	\label{fig:Model}
\end{figure}

In this section, we construct a 3D model of coupled Rashba nanowires that exhibits a HOTI phase with gapless helical hinge states. The Rashba nanowires are taken to be aligned along the $x$ direction, see Fig.~\ref{fig:Wires_uniform}. We now define a unit cell consisting of 8 wires, see Fig.~\ref{fig:Model}. A given unit cell is labeled by two discrete indices $(n,m)$ indicating its position along the $y$ and $z$ direction, respectively. Furthermore, the position of a wire within the unit cell is denoted by three indices $(\tau,\eta,\nu)$, where $\tau\in\{1,\bar{1}\}$ denotes the left/right wire with respect to the $y$ direction, $\eta\in\{1,\bar{1}\}$ the top/bottom two wires with respect to the $z$ direction, and $\nu\in\{1,\bar{1}\}$ the top/bottom wire for a given $\eta$, see again Fig.~\ref{fig:Model}. The electrons in wires with $\eta\nu = 1$ ($\eta\nu = \bar{1}$) are taken to have a positive (negative) effective mass $m^*>0$ ($-m^*<0$), such that the kinetic term describing the uncoupled wires takes the form
\begin{align}\label{eq:Kinetic}
H_0 = &\sum\limits_{n,m} \sum\limits_{\sigma,\tau,\eta,\nu}  \int \mathrm{d}x \,
\Psi^\dagger_{nm\sigma\tau\eta\nu}(x)  \nonumber\\ &\quad\times \left[-\eta\nu \left(\frac{\hbar^2\partial_x^2 }{2m^*}+ \mu \right)\right] \Psi_{nm\sigma\tau\eta\nu}(x),
\end{align}
where $\Psi_{nm\sigma\tau\eta\nu}^\dagger(x)$ [$\Psi_{nm\sigma\tau\eta\nu}(x)$] creates [destroys] an electron with spin $\sigma \in\{1,\bar{1}\}$ at the position $x$ in the $(\tau,\eta,\nu)$ wire of the unit cell $(n,m)$ and $\mu$ denotes the chemical potential. Additionally, we assume that the wires have strong spin-orbit interaction (SOI), which determines the spin quantization axis $z$.  Wires with $\tau\nu = 1$ ($\tau\nu = \bar{1}$) are taken to have SOI of strength $\alpha>0$ ($-\alpha<0$), such that the corresponding term in the Hamiltonian reads
\begin{equation}\label{eq:SOI}
H_\text{SOI} = -i  \alpha\sum_{n,m} \sum_{\sigma,\tau,\eta,\nu} \sigma\tau \nu \int \mathrm{d}x \, \Psi^\dagger_{nm\sigma\tau\eta\nu}  \partial_x \Psi_{nm\sigma\tau\eta\nu}.
\end{equation}
Here and in the following, we suppress the position argument of the field operators for brevity. Combining Eqs.~(\ref{eq:Kinetic}) and (\ref{eq:SOI}) and assuming infinitely long wires for now, we find the spectrum of a single unit cell to consist of 16 branches with energies given by
\begin{eqnarray}\label{eq:Spectrum_continuum}
E_{\sigma\tau\eta\nu} = \eta\nu \left[ \frac{\hbar^2 (k_x + \sigma\tau\eta k_{so})^2}{2m^*} - \tilde{\mu}\right].
\end{eqnarray}
Here, we have introduced the spin-orbit momentum $k_{so} = m^*\alpha/\hbar^2$ and the shifted chemical potential $\tilde{\mu}= \mu +\hbar^2k_{so}^2/2m^*$. As such, the original chemical potential $\mu$ now measures the shift away from the spin-orbit energy $E_{{so}}=\hbar^2k_{so}^2/2m^*$.

We now proceed by coupling neighboring nanowires via different tunneling processes. In the following we set $\mu = 0$, such that the chemical potential lies at the crossing point between spin-up and spin-down branches at $k_x = 0$, see Fig.~\ref{fig:Model}. This choice, together with the particular structure of the unit cell, allows for various momentum-conserving tunneling processes between nearest-neighbor wires.

Let us start by describing the tunneling processes along the $z$ direction. Due to the spatial structure of the unit cell (see Fig.~\ref{fig:Model}), all nearest-neighbor tunneling processes along the $z$ direction conserve the $\tau$-index. First, we account for an intercell tunneling of strength $w$ via
\begin{equation}
H_z = w\sum_{n,m} \sum_{\sigma,\tau} \tau \int \mathrm{d}x \, \Psi^\dagger_{n(m+1)\sigma\tau\bar{1}\bar{1}} \Psi_{nm\sigma\tau11} + \text{H.c.}
\label{eq:Inter_z}
\end{equation}
This term couples neighboring wires in adjacent unit cells and has a sign determined by $\tau$. Furthermore, we introduce an intracell tunneling of strength $w'$ (strength $u$) that couples neighboring wires with the same $\eta$ (with opposite $\eta$) within the same unit cell. These terms read
\begin{align}
H_z' &= w'\sum_{n,m}\sum_{\sigma,\tau,\eta} \tau \int \mathrm{d}x \, \Psi^\dagger_{nm\sigma\tau\eta 1}\Psi_{nm\sigma\tau\eta \bar{1}} + \text{H.c.}, \label{eq:Intra_z} \\
H_z'' &= u\sum_{n,m}\sum_{\sigma,\tau} \tau \int \mathrm{d}x \, \Psi^\dagger_{nm\sigma\tau1\bar{1}} \Psi_{nm\sigma\tau\bar{1}1} + \text{H.c.} \label{eq:u}
\end{align}
Again, the sign of these terms depends on the index $\tau$.
 
Next, we describe the tunneling processes along the $y$ direction. By construction, tunneling terms coupling nearest-neighbor wires along the $y$ direction conserve the $\eta$- and $\nu$-index. First of all, we account for an intercell term with an $\eta$-dependent strength $v_\eta$ via
\begin{equation}
H_y = \sum_{n,m} \sum_{\sigma,\eta,\nu} \int \mathrm{d}x \, v_\eta \, \Psi^\dagger_{(n+1)m\sigma1\eta\nu} \Psi_{nm\sigma\bar{1}\eta\nu} + \text{H.c.} \label{eq:Inter_y}
\end{equation}
This term couples neighboring wires in adjacent unit cells. Similarly, we also introduce an $\eta$-dependent intracell tunneling of strength $v_{\eta}'$ between neighboring wires of the same unit cell:
\begin{equation}
H_y' = \sum_{n,m} \sum_{\sigma,\eta,\nu} \int \mathrm{d}x \, v_\eta' \, \Psi^\dagger_{nm\sigma1\eta\nu} \Psi_{nm\sigma\bar{1}\eta\nu} + \text{H.c.} \label{eq:Intra_y}
\end{equation}
All of the interwire terms given in Eqs.~(\ref{eq:Inter_z})-(\ref{eq:Intra_y}) are illustrated in Fig.~\ref{fig:Model}. Finally, the total Hamiltonian $H$ of our coupled-wires model is given by the sum
\begin{equation}
H = H_0 + H_\text{SOI} + H_y + H_y' + H_z + H_z' + H_z''.\label{eq:HOTI}
\end{equation}
This Hamiltonian is time-reversal symmetric and conserves the $z$-component of the spin since only spin-conserving interwire tunnelings were included. However, once the system enters the HOTI phase (see Sec.~\ref{sec:HOTI} below), also processes in which the spin gets flipped will not change the topological properties of the system as long as the bulk and surface gaps remain open and time-reversal symmetry is maintained.

In the following, all of the interwire hopping amplitudes are considered to be small compared to the spin-orbit energy such that they can be treated as weak perturbations to Eq.~(\ref{eq:Spectrum_continuum}). This allows us to linearize the spectrum of each wire around the respective Fermi points in order to simplify the description of our model. As such, we introduce slowly varying right- and left-moving fields $R_{nm\sigma\tau\eta\nu}(x)$ and $L_{nm\sigma\tau\eta\nu}(x)$ via
\begin{equation}
\Psi_{nm\sigma\tau\eta\nu} =  R_{nm\sigma\tau\eta\nu} \, e^{ik_F^{1\sigma\tau\eta\nu}x} + L_{nm\sigma\tau\eta\nu} \, e^{ik_F^{\bar{1}\sigma\tau\eta\nu}x},
\end{equation}
where $k_F^{r\sigma\tau\eta\nu}$ are the Fermi momenta of the respective branches. Explicitly, these are given by $k_F^{(\eta\nu)(\bar{\tau}\eta)\tau\eta\nu} = 2k_{so}$, $k_F^{(\eta\nu)(\tau\eta)\tau\eta\nu} = k_F^{(\bar{\eta}\nu)(\bar{\tau}\eta)\tau\eta\nu} = 0$, and $k_F^{(\bar{\eta}\nu)(\tau\eta)\tau\eta\nu} = -2k_{so}$. From now on, to simplify our notation, we work in terms of the Hamiltonian density $\mathcal{H}$ defined via $H=\sum_{n,m}\int \mathrm{d}x\,\mathcal{H}(x)$ and give all individual contributions to the Hamiltonian in this form. In terms of right- and left-movers, the effective Hamiltonian of the uncoupled wires is then given by
\begin{align}
\mathcal{H}_0 &= -i\hbar \sum_{\sigma,\tau,\eta,\nu} v_F (R^\dagger_{nm\sigma\tau\eta\nu} \partial_x R_{nm\sigma\tau\eta\nu} \nonumber\\
&\hspace{27mm}- L^\dagger_{nm\sigma\tau\eta\nu}\partial_x L_{nm\sigma\tau\eta\nu})
\end{align}
with the Fermi velocity $v_F = \alpha/\hbar$. In the same way, we can rewrite the tunneling terms in Eqs.~(\ref{eq:Inter_z})-(\ref{eq:Intra_y}) in terms of the new fields. Neglecting rapidly oscillating terms, we obtain
\begin{align}
\mathcal{H}_z =& \sum_{\tau} w\tau( R^\dagger_{n(m+1)\bar{\tau}\tau\bar{1}\bar{1}}L_{nm\bar{\tau}\tau 11} \nonumber\\&\hspace{11mm}+L^\dagger_{n(m+1)\tau\tau\bar{1}\bar{1}} R_{nm\tau\tau11})+\text{H.c.},\\
\mathcal{H}_z' = &\sum_{\sigma,\tau,\eta,\nu}w' \tau R^\dagger_{nm\sigma\tau\eta\nu} L_{nm\sigma\tau\eta\bar{\nu}} + \text{H.c.},\\
\mathcal{H}_z'' =& \sum_{\tau,\nu} u\tau R^\dagger_{nm(\tau\nu)\tau\bar{\nu}\nu}L_{nm(\tau\nu)\tau\nu\bar{\nu}} + \text{H.c.},\\ 
\mathcal{H}_y =& \sum_{\eta,\nu} v_\eta\,(R^\dagger_{(n+1)m\nu1\eta\nu} L_{nm\nu\bar{1}\eta\nu} \nonumber\\&\hspace{11mm}+ L^\dagger_{(n+1)m\nu1\eta\bar{\nu}} R_{nm\nu\bar{1}\eta\bar{\nu}}) + \text{H.c.},\\
\mathcal{H}_y' =&\sum_{\tau,\eta,\nu} v'_\eta R^\dagger_{nm(\tau\nu)\tau\eta\nu}L_{nm(\tau\nu)\bar{\tau}\eta\nu}  + \text{H.c.}
\end{align}

\begin{figure*}[tb]
	\centering
	\includegraphics[width = \textwidth]{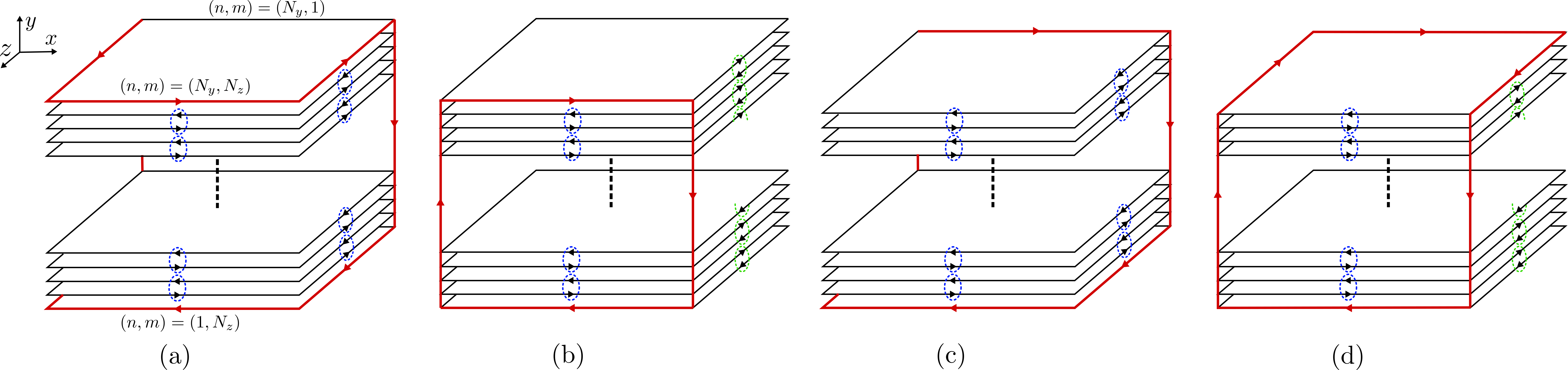}
	\caption{If the interwire tunneling amplitudes along the $y$ direction are set to zero, the 3D coupled-wires model defined in Eq.~(\ref{eq:HOTI}) effectively realizes a stack of uncoupled 2D QSH layers (black rectangles). Small but non-zero $v_\eta$, $v_\eta'$ then lead to the opening of gaps in the $xy$ and $yz$ surfaces by coupling counterpropagating QSH edge states in adjacent layers. Generally, gapless hinge states (red) will be found propagating along some edges of the first and last QSH layer of the stack as well as along hinges where two surfaces with incompatible dimerization patterns (shown pictorially in blue and green) meet. (a) For $v_1>v_{\bar{1}}' \gg v_{\bar{1}},v_1'$, the two $yz$ surfaces are dimerized in a nontrivial way (blue ovals). This matches the pattern of the $xy$ surface at $m=N_z$. (b) For $v_{\bar{1}}'>v_1 \gg v_{\bar{1}},v_1'$, the two $yz$ surfaces are dimerized in a trivial way (green ovals). This matches the pattern of the $xy$ surface at $m=1$. (c) If the last QSH layer of the stack is removed, one of the gapless hinge states is relocated from the $m=N_z$ plane to the $m=1$ plane. Here, we depict the case $v_1>v_{\bar{1}}' \gg v_{\bar{1}},v_1'$. (d) Same as in (c), but with $v_{\bar{1}}'>v_1 \gg v_{\bar{1}},v_1'$. For simplicity, only one spin sector is depicted in all panels.}
	\label{fig:HOTI}
\end{figure*}

\section{Gapless hinge states}
\label{sec:HOTI}
In the following, we focus on the parameter hierarchy $ u,w \gg w'\gg  v_1, v'_{\bar{1}}  \gg v_1', v_{\bar{1}} \geq 0$ and show that, in this regime, our system is a HOTI with a Kramers pair of gapless hinge states propagating along a closed path of hinges of a 3D sample. The required parameter hierarchy could, for example, be achieved by an appropriate spatial arrangement of the wires, as the interwire tunneling amplitudes are expected to decrease with increasing distance between the wires.

Let us start by considering a system of infinite extent along the $x$ direction but a finite number of unit cells $N_y\times N_z$ along the $y$ and $z$ directions, respectively. We first discuss the tunneling terms along the $z$ direction as these are assumed to be dominant. First of all, it is clear that the exterior modes with Fermi momenta at $\pm 2k_{{so}}$ are fully gapped by the $w'$ term since this is the only term affecting these modes. For the interior modes with zero Fermi momentum, on the other hand, the $u$ and $w$ terms compete with the $w'$ term. In the regime we consider, the $u$ and $w$ terms are dominant and couple the interior modes in a pairwise fashion, such that the bulk of the system is fully gapped. However, in a finite system, there are four modes per unit cell in the first ($m=1$) and last ($m=N_z$) layer of wires with respect to the $z$ direction that cannot be paired up, see Fig.~\ref{fig:Model}. Explicitly, these modes are given by the Kramers pairs $R_{n1\bar\tau\tau\bar{1}\bar{1}}$, $L_{n1\tau\tau\bar{1}\bar{1}}$ and $R_{nN_z\tau\tau11}$, $L_{nN_z\bar{\tau}\tau11}$. As a next step,  the tunneling terms along the $y$ direction are added. Here, the hierarchy $v_1,v'_{\bar{1}}  \gg  v_1', v_{\bar{1}}$ is chosen such that the intercell (intracell) tunneling dominates for $\eta=1$ ($\eta=\bar{1}$). By inspection of Fig.~\ref{fig:Model}, it becomes clear that these terms gap out all remaining modes except for the two Kramers pairs $R_{1N_z1111}$, $L_{1N_z\bar{1}111}$ and $R_{N_yN_z\bar{1}\bar{1}11}$, $L_{N_yN_z1\bar{1}11}$, which again cannot be paired up. These remaining gapless states are tightly localized to two hinges of our sample at $(n,m)=(1,N_z)$ and $(n,m)=(N_y,N_z)$ and therefore correspond to the hinge states we are looking for.

Up to now, we have assumed a system that is infinite along the $x$ direction. However, the question remains as to what happens in a finite 3D sample. We start by noting that, in the limit where there is no coupling along the $y$ direction, the system with $u,w\gg w'$ is nothing but a stack of 2D quantum spin Hall (QSH) insulators~\cite{Sagi2015b} stacked along the $y$ direction, see Fig.~\ref{fig:HOTI}. Note that QSH edge states with the same spin projection propagate in opposite directions for layers with opposite $\tau$ indices. By construction, our 3D stack of QSH layers has fully gapped $xz$ surfaces (with the surface gap equal to the QSH gap of the individual layers), while the other surfaces are gapless. To see what happens to these surfaces once the layers are coupled, it is instructive to project the tunneling terms along the $y$ direction onto the gapless edge states of the 2D QSH layers. From Fig.~\ref{fig:Model}, we can read off that the two $xy$ surfaces are gapped in opposite dimerization patterns: the surface at $m=1$ is gapped out trivially by the $v_{\bar{1}}'$ process, while the surface at $m=N_z$ is gapped nontrivially by the $v_1$ process and hosts two gapless hinge states as discussed above. For the $yz$ surfaces, on the other hand, we find that the $v_1$ and $v_{\bar{1}}'$ processes compete. For simplicity, we focus on the case $u=w$ in the following. In this case, we find that the $yz$ surfaces are gapless if $v_1=v_{\bar{1}}'$ and fully gapped otherwise, see App.~\ref{app:yz_surfaces}. The dimerization pattern according to which the $yz$ surfaces are gapped out depends on whether $v_1$ or $v_{\bar{1}}'$ dominates, see Figs.~\ref{fig:HOTI}(a) and (b). Generally, gapless hinge states will be found along some edges of the first and last QSH layer of the stack as well as along hinges where two topologically inequivalent surfaces---i.e., in our case, surfaces with incompatible dimerization patterns---meet, see again Fig.~\ref{fig:HOTI}.

\begin{figure*}[tb]
	\centering
	\includegraphics[width = \textwidth]{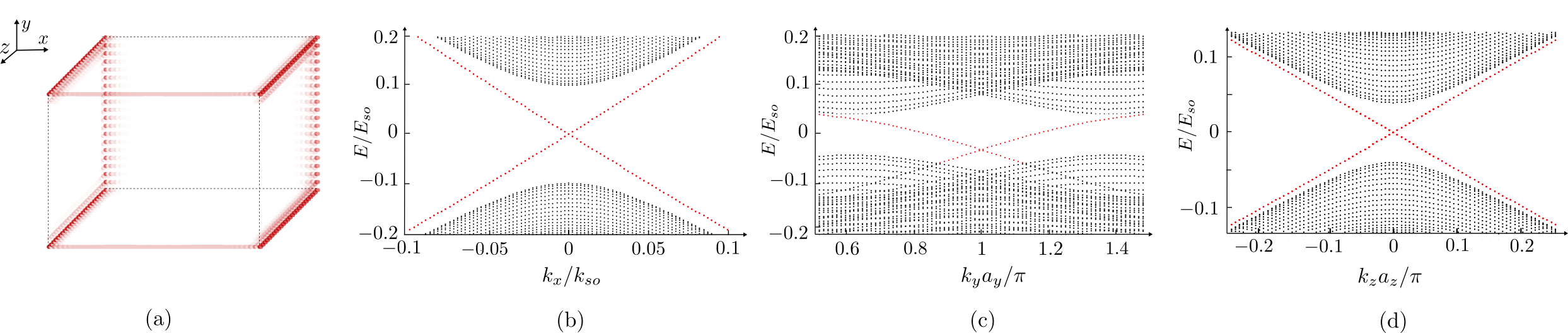}
	\caption{(a) Numerically calculated probability density of the lowest-energy eigenstate of a discretized version of Eq.~(\ref{eq:HOTI}) in a finite sample with $N_y\times N_z=20\times 30$ unit cells and wires of length $\ell =160/k_{{so}}$. Here, we have chosen $v_1>v_{\bar{1}}' \gg v_{\bar{1}},v_1'$, such that we find ourselves in the situation illustrated in Fig.~\ref{fig:HOTI}(a). Indeed, the hinge state (red) follows the expected path around the sample. (b)-(d) Zoom around the Dirac point of the edge state spectrum in dependence on $k_x$, $k_y$, and $k_z$ for a system that is taken to be infinite along the $x$, $y$, or $z$ direction, respectively. In all three cases, the energies of the helical edge states (red dots) are clearly visible inside the gap. Each branch of the edge-state spectrum is twofold (fourfold) degenerate for edge states propagating along the $x$ and $y$ ($z$) direction in agreement with the results shown in the panel (a).  The numerical parameters are $u=w\approx0.78E_{{so}}$, $w'\approx0.31E_{{so}}$, $v_1\approx0.34E_{{so}}$, $v_{\bar{1}}'\approx0.19E_{{so}}$, and $v_{\bar{1}}=v_1'=0$ for all panels.}
	\label{fig:PD_tauz}
\end{figure*}

Furthermore, we note that also the boundary termination of the sample can influence the path of the hinge states~\cite{Yoon2020,Mook2021}. Indeed, for the construction presented so far, both pairs of gapless hinge states propagating along the $x$ direction lie in the plane at $m=N_z$. This can be changed if the surface termination of the sample is adjusted. If, for example, the last QSH layer of the stack (i.e., the $\tau=\bar{1}$ part of all unit cells in the plane $n=N_y$) is removed from the sample, one pair of hinge states is relocated from the $m=N_z$ plane to the $m=1$ plane, see Figs.~\ref{fig:HOTI}(c) and (d). 

All of these findings can be checked numerically by exact diagonalization. In the following, we focus on the situation shown in Fig.~\ref{fig:HOTI}(a), but the other cases can be treated in a similar way. The probability density of the lowest-energy state in a finite 3D sample reveals the presence of tightly localized hinge states following the expected path of hinges, see Fig.~\ref{fig:PD_tauz}(a). In Figs.~\ref{fig:PD_tauz}(b)--(d), a zoom around the Dirac point of the edge state spectrum in dependence on $k_x$, $k_y$, and $k_z$ for a system taken to be infinite along the $x$, $y$, or $z$ direction, respectively, confirms the presence of helical edge states with in-gap energies. Note that the energy gaps in the three semi-infinite geometries generally differ in size since different surfaces are gapped by different mechanisms. This also explains the different localization lengths for hinge states propagating along the $x$, $y$, or $z$ direction. We also note that while our analytical arguments were based on the strictly perturbative regime $u,w \gg w'\gg  v_1, v'_{\bar{1}}  \gg v_1', v_{\bar{1}} \geq 0$, the topological phase is actually found for a broader range of parameters. In particular, the requirement of  $w'\gg  v_1, v'_{\bar{1}}$ can be relaxed, see also App.~\ref{app:yz_surfaces}. Furthermore, we have checked numerically that the hinge states are robust against potential disorder as long as the bulk and surface gaps remain open, see App.~\ref{app:disorder}.

\section{Fractional second-order topological insulator}
\label{sec:FractionalHOTI}
\begin{figure}[!b]
	\centering
	\includegraphics[width=0.75\columnwidth]{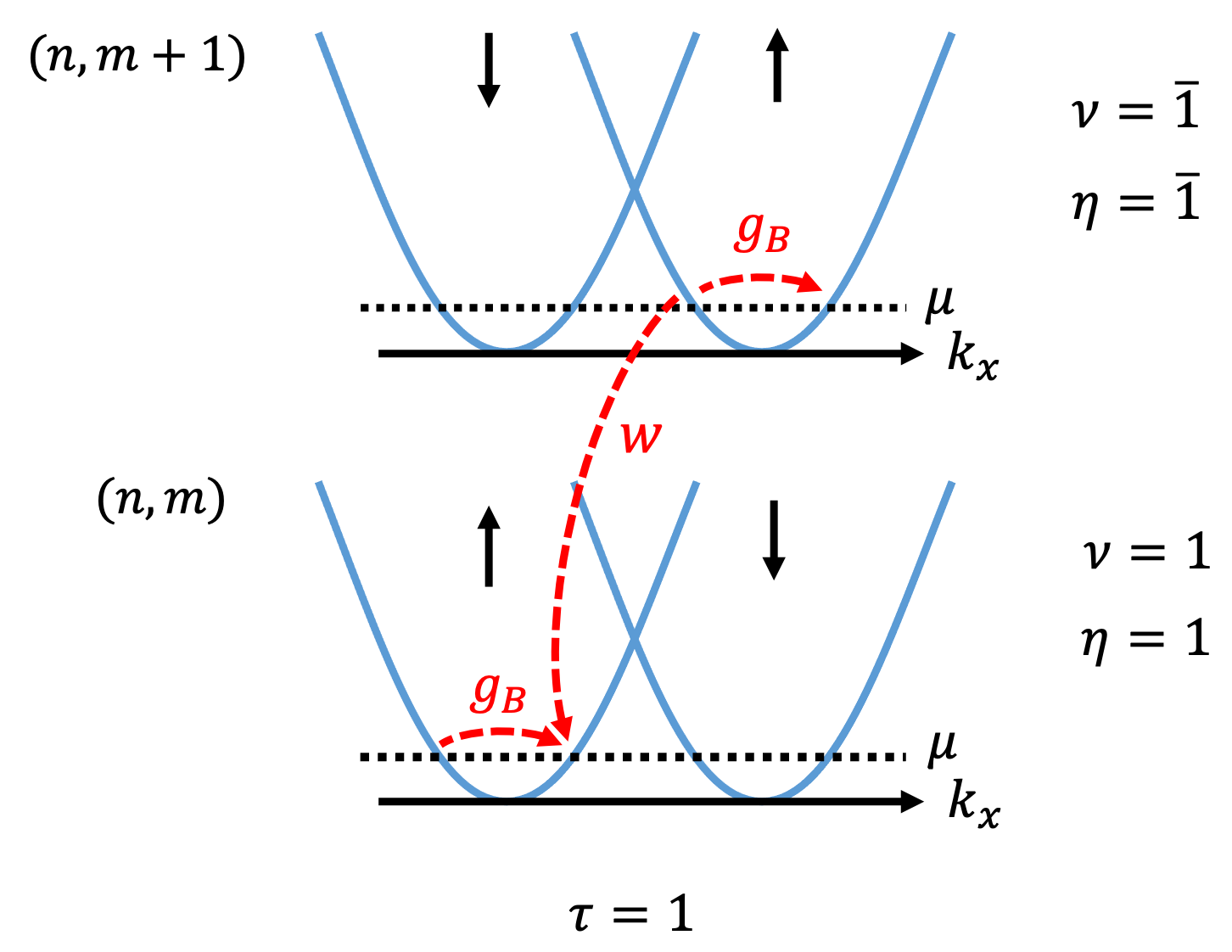}
	\caption{Pictorial representation of an interwire tunneling term in Eq.~(\ref{eq:w_dressed}) for $p = 3$. The chemical potential is tuned to $\mu=-8E_{{so}}/9$, such that the momentum mismatch of the interwire tunneling can be exactly compensated by two backscattering terms of strength $g_B$.}
	\label{fig:Model_interactions}
\end{figure}

In this section, we show how the HOTI found in Sec.~\ref{sec:HOTI} can be promoted to a more exotic, \emph{fractional} HOTI with gapless helical hinge states that carry only a fraction of the electronic charge $e$. For this, we start by shifting the chemical potential to a fractional value
\begin{equation}
\mu = (-1+1/p^2)E_{so}
\end{equation}
for an odd integer $p$. The new Fermi momenta of the uncoupled wires are now given by $k_F^{(\eta\nu)(\bar{\tau}\eta)\tau\eta\nu} = k_{so}(1+1/p)$, $k_F^{(\bar{\eta}\nu)(\tau\eta)\tau\eta\nu} = -k_{so}(1+1/p)$, $k_F^{(\bar{\eta}\nu)(\bar{\tau}\eta)\tau\eta\nu} = k_{so}(1-1/p)$, and $k_F^{(\eta\nu)(\tau\eta)\tau\eta\nu} = -k_{so}(1-1/p)$. For $p = 1$, we retrieve the case discussed in Sec.~\ref{sec:HOTI}. For $p > 1$, on the other hand, the interwire tunneling terms given in Eqs.~(\ref{eq:Inter_z}) and (\ref{eq:u})-(\ref{eq:Intra_y}) do no longer conserve momentum and can therefore not lead to the opening of gaps. However, new momentum-conserving processes can be constructed by taking into account single-electron backscattering processes originating from electron-electron interactions. For the interwire tunneling terms along the $z$ direction given in Eqs.~(\ref{eq:Inter_z}) and (\ref{eq:u}), these new momentum-conserving multi-electron processes read
\begin{widetext}
\begin{align}
\tilde{\mathcal{H}}_z &= \tilde{w}\sum_\tau \tau\big[(R^\dagger_{n(m+1)\bar{\tau}\tau\bar{1}\bar{1}}L_{n(m+1)\bar\tau\tau\bar{1}\bar{1}})^q(R^\dagger_{n(m+1)\bar\tau\tau\bar{1}\bar{1}}L_{nm\bar\tau\tau 11})(R^\dagger_{nm\bar\tau\tau 11}L_{nm\bar\tau\tau 11})^q \nonumber\\&\hspace{18mm}+ (L^\dagger_{n(m+1)\tau\tau\bar{1}\bar{1}}R_{n(m+1)\tau\tau\bar{1}\bar{1}})^q(L^\dagger_{n(m+1)\tau\tau\bar{1}\bar{1}} R_{nm\tau\tau 11})(L^\dagger_{nm\tau\tau 11}R_{nm\tau\tau 11})^q\big]+ \text{H.c.},\label{eq:w_dressed}\\
\tilde{\mathcal{H}}_z'' &=  \tilde{u}\sum_{\tau,\nu}\tau\,  (R^\dagger_{nm(\tau\nu)\tau\bar{\nu}\nu}L_{nm(\tau\nu)\tau\bar{\nu}\nu})^q (R^\dagger_{nm(\tau\nu)\tau\bar{\nu}\nu}L_{nm(\tau\nu)\tau\nu\bar{\nu}})(R^\dagger_{nm(\tau\nu)\tau\nu\bar{\nu}}L_{nm(\tau\nu)\tau\nu\bar{\nu}})^q + \text{H.c.},\label{eq:u_dressed}
\end{align}
where we have defined $q=(p-1)/2$. The amplitudes of the above terms are given by $\tilde{w} \propto w \, g^{p-1}_B$ and $\tilde{u} \propto u \, g^{p-1}_B$, where $g_B$ denotes the strength of a single-electron backscattering process. In Fig.~\ref{fig:Model_interactions}, we pictorially represent a process in Eq.~(\ref{eq:w_dressed}) for the case $p = 3$. In the following, let us assume that the above terms are relevant in the renormalization group (RG) sense and that they are the dominant interwire tunneling terms in our model. This can always be achieved if their bare coupling constants are sufficiently large or if their scaling dimensions are the lowest ones among all possible competing terms. Next, we consider the $w'$ term given in Eq.~(\ref{eq:Intra_z}). While this term conserves momentum for all $p$, it does not commute with the terms given in Eqs.~(\ref{eq:w_dressed}) and (\ref{eq:u_dressed}) and can therefore not order simultaneously. To lowest order in the interaction, the term that commutes with Eqs.~(\ref{eq:w_dressed}) and (\ref{eq:u_dressed}) is again given by the `dressed' term
\begin{align}
\tilde{\mathcal{H}}_z' &= \tilde{w}' \sum_{\sigma,\tau,\eta,\nu}\tau\, (R^\dagger_{nm\sigma\tau\eta\nu}L_{nm\sigma\tau\eta\nu})^q(R^\dagger_{nm\sigma\tau\eta\nu} L_{nm\sigma\tau\eta\bar{\nu}})(R^\dagger_{nm\sigma\tau\eta\bar{\nu}}L_{nm\sigma\tau\eta\bar{\nu}})^q+ \text{H.c.},
\end{align}
\end{widetext}
where again $\tilde{w}' \propto w' \, g^{p-1}_B$. We assume that this term is relevant in the RG sense but weak compared to the $u$ and $w$ terms discussed above, such that we have $\tilde{u},\tilde{w}\gg\tilde{w}'$.

To gain further insight into the interacting model, we now switch to a bosonized language~\cite{Giamarchi2004} by writing
\begin{eqnarray}
R_{nm\sigma\tau\eta\nu}(x) &\propto e^{i\phi_{1nm\sigma\tau\eta\nu}(x)}, \nonumber \\ L_{nm\sigma\tau\eta\nu}(x) &\propto e^{i\phi_{\bar{1}nm\sigma\tau\eta\nu}(x)},
\end{eqnarray}
where the bosonic fields $\phi_{rnm\sigma\tau\eta\nu}$ with $r\in \{1,\bar{1}\}$ satisfy the standard non-local commutation relations
\begin{align} \label{eq:commutation}
&[\phi_{rnm\sigma\tau\eta\nu}(x), \phi_{r'n'm'\sigma'\tau'\eta'\nu'}(x')] \nonumber\\ 
&=ir\pi\delta_{nn'}\delta_{mm'}\delta_{rr'}\delta_{\sigma\sigma'}\delta_{\tau\tau'}\delta_{\eta\eta'}\delta_{\nu\nu'}\,\text{sgn}(x - x').
\end{align}
This relation guarantees that fermionic fields of the same species satisfy the correct anticommutation relations. For fermionic fields of different species, the anticommutation relations need to be enforced by a proper choice of Klein factors, which we will not include here explicitly~\cite{Teo2014}. To proceed, it will be useful to introduce a new set of bosonic fields defined as
\begin{align}
\chi_{rnm\sigma\tau\eta\nu} = \frac{p+1}{2} \phi_{rnm\sigma\tau\eta\nu} - \frac{p-1}{2} \phi_{\bar{r}nm\sigma\tau\eta\nu}.
\end{align}
The new fields obey the commutation relations
\begin{align}
&[\chi_{rnm\sigma\tau\eta\nu}(x), \chi_{r'n'm'\sigma'\tau'\eta'\nu'}(x')] \nonumber\\
&=ir\pi p\delta_{nn'}\delta_{mm'}\delta_{rr'}\delta_{\sigma\sigma'}\delta_{\tau\tau'}\delta_{\eta\eta'}\delta_{\nu\nu'}\,\text{sgn}(x - x').\label{eq:commutation_fractional}
\end{align}
In terms of these new fields, the interwire tunneling terms along the $z$ direction then take the relatively simple form
\begin{align}
\tilde{\mathcal{H}}_z& \propto \tilde w \sum_\tau\tau\big[ \cos{(\chi_{\bar{1}n(m+1)\tau\tau \bar{1}\bar{1}} - \chi_{1nm\tau\tau 11})} \nonumber \\&\hspace{17mm}+ \cos{(\chi_{1n(m+1)\bar\tau\tau\bar{1}\bar{1}} - \chi_{\bar{1}nm\bar\tau\tau11})}\big],\label{eq:Hz_cos}\\
\tilde{\mathcal{H}}_z' &\propto\tilde{w}' \sum_{\sigma,\tau,\eta,\nu} \tau \cos{(\chi_{1nm\sigma\tau\eta\nu} - \chi_{\bar{1}nm\sigma \tau\eta\bar{\nu}})},\\
\tilde{\mathcal{H}}_z'' &\propto \tilde{u}\sum_{\nu,\tau} \tau \cos{(\chi_{1nm(\tau\nu)\tau\bar{\nu}\nu} - \chi_{\bar{1}nm(\tau\nu)\tau\nu\bar{\nu}})}.\label{eq:Hu_cos}
\end{align}
Terms of this form frequently appear in the context of coupled-wires models and have been thoroughly studied in previous works~\cite{Kane2002,Teo2014,Klinovaja2013,Klinovaja2014a,Santos2015,Klinovaja2014b,Klinovaja2015,Sagi2014,Sagi2015a,Sagi2015b,Meng2015}. In the regime of strong coupling, which is the regime of interest here, the arguments of the cosines get `pinned' to constant values in order to minimize the corresponding terms in the Hamiltonian. It then becomes clear that Eqs.~(\ref{eq:Hz_cos})-(\ref{eq:Hu_cos}) lead to a fully gapped bulk since all fields are pinned pairwise. In a finite sample, however, the fields $\chi_{1n1\bar\tau\tau\bar{1}\bar{1}}$, $\chi_{\bar{1}n1\tau\tau\bar{1}\bar{1}}$ and $\chi_{1nN_z\tau\tau11}$, $\chi_{\bar{1}nN_z\bar{\tau}\tau11}$ remain gapless.

Similarly to the non-interacting case, we now include interwire tunneling processes along the $y$ direction to gap out all remaining fields except a single pair of helical hinge states. Again, the terms given in Eqs.~(\ref{eq:Inter_y}) and (\ref{eq:Intra_y}) have to be combined with backscattering processes to ensure momentum conservation. For the sake of brevity, we directly write the resulting multi-electron processes in terms of bosonic fields and defer their fermionic expressions to App.~\ref{app:dressed_y}. To lowest order in the interaction, the momentum-conserving tunneling terms are given by
\begin{align}
\tilde{\mathcal{H}}_y &\propto \sum_{\eta,\nu}  \tilde{v}_\eta\big[ \cos{(\chi_{1(n+1)m\nu1\eta\nu} - \chi_{\bar{1}nm\nu\bar{1}\eta\nu})} \nonumber \\&\hspace{15mm}+ \cos{(\chi_{\bar{1}(n+1)m\nu1\eta\bar{\nu}} - \chi_{1nm\nu\bar{1}\eta\bar{\nu}})}\big],\\
\tilde{\mathcal{H}}_y'&\propto \sum_{\tau,\eta,\nu} {\tilde{v}'}_\eta\cos{(\chi_{1nm(\tau\nu)\tau\eta\nu} - \chi_{\bar{1}nm(\tau\nu)\bar{\tau}\eta\nu})},
\end{align}
where again $\tilde{v}_\eta \propto v_\eta \, g^{p-1}_B$ and $\tilde{v}_\eta' \propto v_\eta' \, g^{p-1}_B$. If these terms are relevant but weaker than the tunneling terms along the $z$ direction discussed previously, we find that the bulk of the system is fully gapped, while for a finite system of $N_y \times N_z$ unit cells the fields $\chi_{11N_z1111}$, $\chi_{\bar{1}1N_z\bar{1}111}$, $\chi_{1N_yN_z\bar{1}\bar{1}11}$, and $\chi_{\bar{1}N_yN_z1\bar{1}11}$ stay gapless. This reveals the presence of two Kramers pairs of gapless hinge states propagating along the $x$ direction. Importantly, these gapless edge states now carry a fractional charge $e/p$~\cite{Teo2014}, see App.~\ref{app:FQHE} for a brief review.

Indeed, we can understand the emergence of the fractional HOTI phase in the same way as is shown in Fig.~\ref{fig:HOTI}, with the only difference that the QSH layers are now replaced by \emph{fractional} QSH layers~\cite{Sagi2015b}. In the same way as before, gapless hinge states propagate along certain edges of the first and last layer of the stack as well as along hinges where surfaces with incompatible dimerization patterns meet. As such, the path of the fractional hinge states can again be controlled by adjusting the interwire tunneling amplitudes as well as the surface termination of the sample.

\section{Conclusions}
\label{sec:conclusions}

We have constructed a 3D model of coupled Rashba nanowires that can realize a HOTI phase with a Kramers pair of gapless helical hinge states propagating along a closed path of hinges of a finite sample. The specific choice of interwire tunneling amplitudes and surface termination allow for control over the path that the gapless hinge states take. Moreover, the coupled-wires approach allows us to incorporate strong electron-electron interactions into our description. For sufficiently strong interactions, we have shown that the system can enter a fractional HOTI phase with gapless hinge states that carry only a fraction of the electronic charge $e$.

The emergence of hinge states can be intuitively understood by viewing our model as a stack of (fractional) QSH layers that are coupled in a nontrivial way such that different surfaces are gapped out in different, incompatible dimerization patterns. We note that while we have focused on the time-reversal invariant case, we can straightforwardly obtain a formal description the time-reversal broken case with a (fractional) chiral hinge state by focusing on just one spin sector of our model. More physically, one could also think of modeling a (fractional) chiral HOTI by starting from a coupled-wires description of a stack of (fractional) quantum Hall layers and then coupling adjacent layers in suitable dimerization patterns in a similar way as in this work.

Finally, while coupled-wires models are mainly a theoretical tool to analytically construct and describe exotic, strongly interacting phases, we note that some aspects of our construction---in particular the picture of coupled QSH layers---are closely related to recent experimental work on 3D HOTIs in layered systems~\cite{Noguchi2021,Shumiya2022}.

\acknowledgments
This work was supported by the Swiss National Science Foundation and NCCR QSIT. This project received funding from the European Union’s Horizon 2020 research and innovation program (ERC Starting Grant, grant agreement No 757725).

\appendix

\begin{widetext}
\section{Gap-opening terms in the $yz$ plane}
\label{app:yz_surfaces}

In this Appendix, we discuss the details of how the interwire tunneling terms along the $y$ direction lead to the HOTI phase discussed in the main text. When only the interwire tunneling terms along the $z$ direction are taken into account, the system corresponds to a stack of uncoupled 2D QSH layers stacked along the $y$ direction. By construction, it is clear that the $xy$ surfaces are fully gapped by the hopping terms along the $y$ direction, see Fig.~\ref{fig:Model}. The question remains what happens to the $yz$ surfaces. To see this, we can look at the explicit expressions for the uncoupled QSH edge states in a system that is assumed to be infinite along the $z$ direction, such that $k_z$ is a good quantum number. For simplicity, we focus on the case $u=w$. At $k_z=0$, the gapless QSH edge states in the $n$th layer can be labeled by a fixed spin $\sigma$ and a fixed $\tau$, while their $\eta$- and $\nu$-components are given by four-component wave functions $\phi_{n\tau\sigma}(x)$. These are found to be
\begin{align}
\phi_{n\tau 1}(x)&=\frac{1}{\sqrt{\mathcal{N}}}\left[\begin{pmatrix}i\\-1\\-i\\1\end{pmatrix}e^{-x/\xi_1}
+\begin{pmatrix}-ie^{2ik_{so}x}\\e^{2ik_{so}x}\\ie^{-2ik_{so}x}\\-e^{-2ik_{so}x}\end{pmatrix}e^{-x/\xi_2}\right],\\
\phi_{n\tau \bar{1}}(x)&=\frac{1}{\sqrt{\mathcal{N}}}\left[\begin{pmatrix}-i\\-1\\i\\1\end{pmatrix}e^{-x/\xi_1}
+\begin{pmatrix}ie^{-2ik_{so}x}\\e^{-2ik_{so}x}\\-ie^{2ik_{so}x}\\-e^{2ik_{so}x}\end{pmatrix}e^{-x/\xi_2}\right],
\end{align}
where $\mathcal{N}$ is a normalization constant. Here, we have defined $\xi_1=\alpha/(u-w')$ and $\xi_2=\alpha/w'$. %The wave functions for $\tau=\bar{1}$ can be obtained from the above by reversing the role of spin up and spin down.
We can now calculate the projection of the $v_1$ and $v_{\bar{1}}'$ terms onto these edge states. The $v_1$ term couples edge states in adjacent unit cells via the matrix element
\begin{equation}
v_1\left\langle \phi_{(n+1)1\sigma}(x)\Big|\frac{\mathbb{1}+\eta_z}{2}\Big|\phi_{n\bar{1}\sigma}(x)\right\rangle=\frac{v_1}{2},
\end{equation}
where $\eta_z$ is a Pauli matrix acting on the $\eta$ subspace (see Fig.~\ref{fig:Model} for an illustration). The $v_{\bar{1}}'$ term, on the other hand, couples edge states in the same unit cell via the matrix element
\begin{equation}
v_{\bar{1}}'\left\langle \phi_{n1\sigma}(x)\Big|\frac{\mathbb{1}-\eta_z}{2}\Big|\phi_{n\bar{1}\sigma}(x)\right\rangle=\frac{v_{\bar{1}}'}{2}.
\end{equation}
We now see that, at $v_1=v_{\bar{1}}'$, the gaps induced by the two terms are equal and the $yz$ surfaces are thus gapless. Otherwise, the $yz$ surfaces will be fully gapped and the corresponding dimerization pattern will be set by whichever one of the two terms dominates. Additionally, we see that, due to the spatial structure of the unit cell, the gaps opened by the tunneling  terms along the $y$ direction are suppressed by an additional factor of $1/2$. Generally, the strict requirement $u,w \gg w'\gg  v_1, v'_{\bar{1}}$ can be relaxed as long as the 2D bulk gap of the individual QSH layers [given by $E_\mathrm{gap}=\mathrm{min}(u-w',w')$] is not closed by the tunneling terms along the $y$ direction.

\section{Stability against potential disorder}
\label{app:disorder}

In Fig.~\ref{fig:disorder}, we present numerical data demonstrating that the hinge states found in the main text are robust against potential disorder. To model the disorder, we allow for random fluctuations of the chemical potential following a normal distribution with zero mean value and a standard deviation of $\sigma_\mu$. Figure~\ref{fig:disorder}(b) shows that the hinge states are robust against potential disorder even if $\sigma_\mu$ significantly exceeds the minimum surface gap $\Delta$ ($\Delta\approx 0.13E_{so}$ for the parameters used in Fig.~\ref{fig:PD_tauz}). However, if $\sigma_\mu$ gets even larger, the surface gap closes and the hinge states tend to get localized, see Fig.~\ref{fig:disorder}(c).

	\begin{figure}[bt]
	\centering
	\includegraphics[width=0.9\columnwidth]{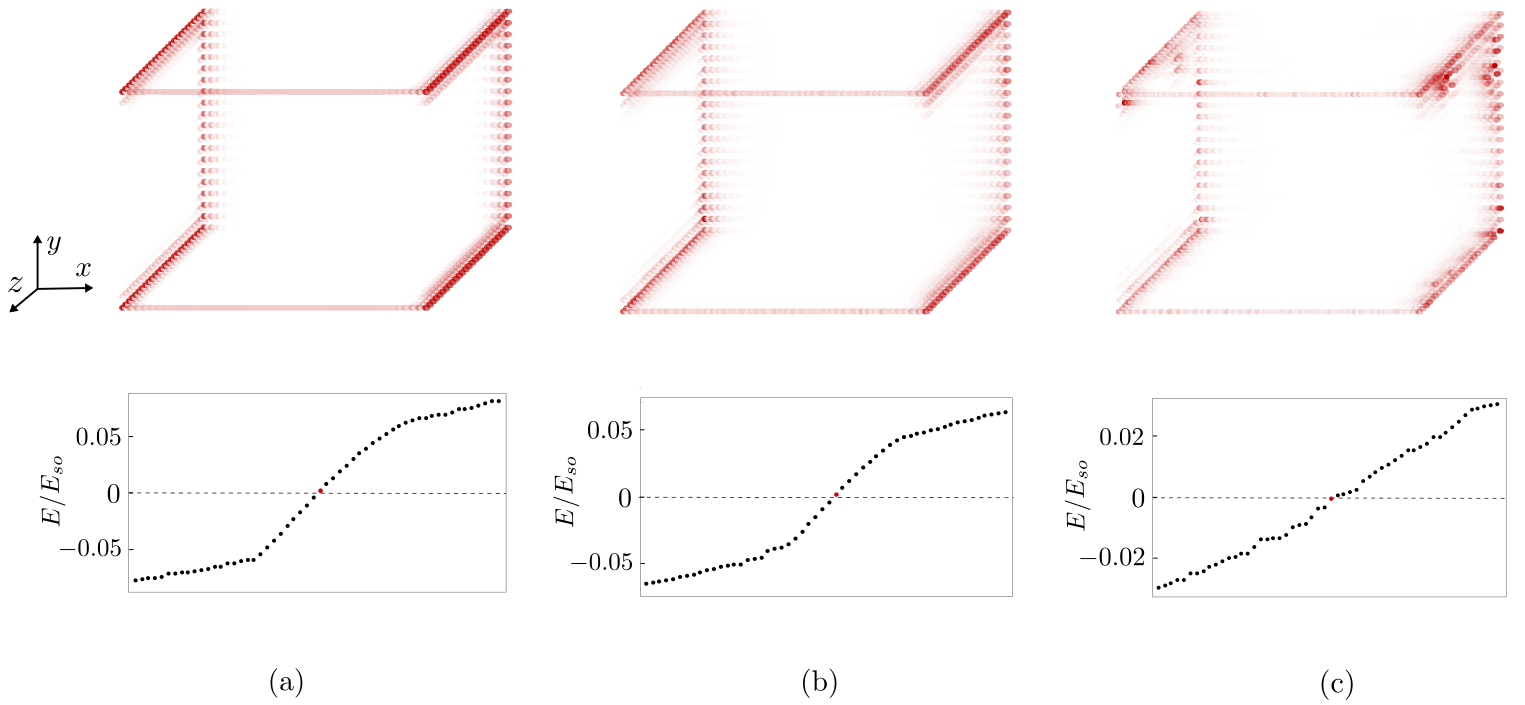}
	\caption{Numerically calculated probability density of the lowest-energy eigenstate (top row) and low-energy spectrum (bottom row) of a discretized version of Eq.~(\ref{eq:HOTI}) with additional potential disorder drawn from a normal distribution with zero mean value and standard deviation $\sigma_\mu$. (a) $\sigma_\mu=0$. (b) $\sigma_\mu\approx0.31E_{{so}}$. (c) $\sigma_\mu\approx0.55 E_{so}$. We find that the hinge states are robust against potential disorder even if $\sigma_\mu$ exceeds the minimum surface gap $\Delta\approx 0.13E_{{so}}$, see panel (b). However, if $\sigma_\mu$ gets too large, the surface gap closes and the hinge states tend to get localized, see panel (c). The parameters are the same as in Fig.~\ref{fig:PD_tauz}.}
	\label{fig:disorder}
\end{figure}

\section{Dressed interwire tunneling terms along the $y$ direction}
\label{app:dressed_y}

For completeness, this Appendix gives the full fermionic form of the dressed interwire tunneling terms along the $y$ direction. These read:
\begin{align}
\tilde{\mathcal{H}}_y&=\sum_{\eta,\nu} \tilde{v}_\eta\,\big[(R^\dagger_{(n+1)m\nu1\eta\nu} L_{(n+1)m\nu1\eta\nu})^q(R^\dagger_{(n+1)m\nu1\eta\nu} L_{nm\nu\bar{1}\eta\nu})(R^\dagger_{nm\nu\bar{1}\eta\nu} L_{nm\nu\bar{1}\eta\nu}) ^q \nonumber\\&\hspace{10mm}+ (L^\dagger_{(n+1)m\nu1\eta\bar{\nu}} R_{(n+1)m\nu1\eta\bar{\nu}})^q(L^\dagger_{(n+1)m\nu1\eta\bar{\nu}} R_{nm\nu\bar{1}\eta\bar{\nu}})(L^\dagger_{nm\nu\bar{1}\eta\bar{\nu}} R_{nm\nu\bar{1}\eta\bar{\nu}})^q\big] + \text{H.c.},\\
\tilde{\mathcal{H}}_y'&=\sum_{\tau,\eta,\nu} \tilde{v}'_\eta \, (R^\dagger_{nm(\tau\nu)\tau\eta\nu}L_{nm(\tau\nu)\tau\eta\nu})^q(R^\dagger_{nm(\tau\nu)\tau\eta\nu}L_{nm(\tau\nu)\bar{\tau}\eta\nu})(R^\dagger_{nm(\tau\nu)\bar{\tau}\eta\nu}L_{nm(\tau\nu)\bar{\tau}\eta\nu})^q + \text{H.c.}
\end{align}

\section{Fractionally charged edge and hinge states}
\label{app:FQHE}

In this Appendix, we briefly review how we can identify fractionally charged edge or hinge states in the framework of a coupled-wire construction. For this, we consider a simplified model that consists only of a single layer (with a fixed unit cell index $n=n_0$ and a fixed layer index $\tau=1$) and a single spin sector (with a fixed spin label $\sigma=1$) of the full model discussed in the main text. This simplified model then hosts a single spin-polarized fractionally charged edge state and effectively reproduces the coupled-wire construction of the fractional quantum Hall effect as discussed in Ref.~\cite{Teo2014}. While we choose to focus on a simplified model for pedagogical reasons, all the arguments presented in this Appendix straightforwardly carry over to the full model with fractionally charged hinge states.

The Hamiltonian of our simplified model is given by Eqs.~(\ref{eq:Hz_cos})--(\ref{eq:Hu_cos}) in the main text, with the only difference that we drop the $n$, $\tau$, and $\sigma$ indices as these are kept fixed throughout this Appendix:
\begin{align}
\bar{\mathcal{H}}_{z}&\propto \tilde w  \cos{(\chi_{\bar{1}(m+1)\bar{1}\bar{1}} - \chi_{1m11})},\label{eq:HzApp}\\
\bar{\mathcal{H}}'_{z} &\propto\tilde{w}' \sum_{\eta,\nu} \cos{(\chi_{1m\eta\nu} - \chi_{\bar{1}m\eta\bar{\nu}})},\\
\bar{\mathcal{H}}''_{z}&\propto \tilde{u}\sum_{\nu} \cos{(\chi_{1m\bar{\nu}\nu} - \chi_{\bar{1}m\nu\bar{\nu}})}.\label{eq:HuApp}
\end{align}
Here, we use the shorthand notations $\phi_{rn_0m11\eta\nu}\equiv \phi_{rm\eta\nu}$, $\chi_{rn_0m11\eta\nu}\equiv \chi_{rm\eta\nu}$. Let us first look at the bulk excitations in an infinite system. As an example, we focus on $\bar{\mathcal{H}}_{z}$. In the bulk, $\bar{\mathcal{H}}_{z}$ leads to a pairwise pinning of the bosonic fields $\chi_{\bar{1}(m+1)\bar{1}\bar{1}} - \chi_{1m11}=\pi\ \mathrm{mod}\ 2\pi$ in order to minimize the ground state energy. An elementary excitation then corresponds to a $2\pi$-kink in the argument of one of the cosine terms for some $m=m_0$. We assume that an isolated kink occurs around some position $x=x_0$, such that $[\chi_{\bar{1}(m_0+1)\bar{1}\bar{1}}(x_2) - \chi_{1m_011}(x_2)]-[\chi_{\bar{1}(m_0+1)\bar{1}\bar{1}}(x_1) - \chi_{1m_011}(x_1)]=\pm 2\pi$ for $x_1\ll x_0\ll x_2$. In order to discuss the charge associated with this kink, it is convenient to introduce an alternative set of fields
\begin{align}
\varphi_{m\eta\nu}&=(\phi_{\bar{1}m\eta\nu}-\phi_{1m\eta\nu})/2,\\
\theta_{m\eta\nu}&=(\phi_{\bar{1}m\eta\nu}+\phi_{1m\eta\nu})/2.
\end{align}
The $\varphi_{m\eta\nu}$ fields can be related to charge densities as $\rho_{m\eta\nu}(x)=-e\,\partial_x\varphi_{m\eta\nu}(x)/\pi$~\cite{Giamarchi2004}. By summing over all pinned fields, it is then straightforward to show that a $2\pi$-kink in the combination $\chi_{\bar{1}(m_0+1)\bar{1}\bar{1}} - \chi_{1m_011}$ translates to a $(\pi/p)$-kink in the total sum $\Phi(x)=\sum_{m,\eta,\nu}\varphi_{m\eta\nu}(x)$. An elementary bulk excitation thus carries a charge
\begin{equation}
q=-\frac{e}{\pi}\int_{x_1}^{x_2} dx\, \partial_x\Phi(x)=-\frac{e}{\pi}\left[\Phi(x_2)-\Phi(x_1)\right]= \pm e/p.
\end{equation}
For the example term discussed above, the operators that introduce $2\pi$-kinks are $e^{i\chi_{\bar{1}(m_0+1) \bar{1}\bar{1}}/p}$ and $e^{i\chi_{1m_011}/p}$, see Eq.~(\ref{eq:commutation_fractional}). Note that these operators are not physical in the sense that they cannot be built out of physical fermion operators. Only the tunneling of a fractional charge from one position to another is a physical operator. To see this, consider e.g. a process of the form $e^{i\chi_{\bar{1}m_0 11}/p}e^{-i\chi_{1m_011}/p}=e^{i\phi_{\bar{1}m_011}}e^{-i\phi_{1m_011}}$. This operator moves a kink in $\chi_{\bar{1}(m_0+1)\bar{1}\bar{1}} - \chi_{1m_011}$ (pinned by $\bar{\mathcal{H}}_{z}$) to a kink in $\chi_{1m_01\bar{1}} - \chi_{\bar{1}m_011}$ (pinned by $\bar{\mathcal{H}}'_{z}$). At the same time, in terms of the original free fermions, this process simply corresponds to the backscattering of a single electron.

Finally, in a finite system, we can build a string of quasiparticle tunneling operators that shuttles a fractional charge of $e/p$ from one edge all the way across the bulk to the other edge. As such, the elementary gapless excitations of the edge carry charge $e/p$ like the gapped bulk excitations.
\end{widetext}

\end{document}